%% file: main.tex
\documentclass[sigconf]{acmart}

\def\BibTeX{{\rm B\kern-.05em{\sc i\kern-.025em b}\kern-.08emT\kern-.1667em\lower.7ex\hbox{E}\kern-.125emX}}
\usepackage{booktabs} 
\usepackage{graphicx}
\usepackage{balance}  
\usepackage{url}
\usepackage{hyperref}
\usepackage{amssymb}
\usepackage[ruled,vlined,linesnumbered]{algorithm2e}
\usepackage{graphicx}
\usepackage{epstopdf}
\usepackage{amsmath,epsfig,amsfonts,amssymb,graphicx}
\usepackage{subfigure}
\usepackage{array}
\usepackage{makecell}
\setcopyright{rightsretained}
\usepackage{fancyhdr}
\usepackage{color}
\usepackage{algorithmic}
\usepackage{textcomp}
\usepackage{diagbox}
\usepackage{booktabs}
\usepackage{multirow}

\begin{document}
\copyrightyear{2018}
\acmYear{2018}
\setcopyright{acmlicensed}
\acmConference[Woodstock '18]{Woodstock '18: ACM Symposium on Neural Gaze Detection}{June 03--05, 2018}{Woodstock, NY}
\acmPrice{15.00}
\acmDOI{10.1145/1122445.1122456}
\acmISBN{978-1-4503-9999-9/18/06}


%

%

%

%
\title{Deep Cross Networks with Aesthetic Preference for Cross-domain Recommendation}
%


\author{Jian Liu}
\affiliation{%
  \institution{Soochow University}
  \city{Suzhou}
  \country{China}}
\email{jliu2013@stu.suda.edu.cn}

\author{Pengpeng Zhao}
\affiliation{%
  \institution{Soochow University}
  \city{Suzhou}
  \country{China}}
\email{ppzhao.suda.edu.cn}

\author{Yanchi Liu}
\affiliation{%
  \institution{Rutgers University}
  \city{New Jersey}
  \country{USA}}
\email{yanchi.liu@rutgers.edu}

\author{Victor S. Sheng}
\affiliation{%
  \institution{University of Central Arkansas}
  \city{Conway}
  \country{USA}}
\email{ssheng@uca.edu}

\author{Fuzhen Zhuang}
\affiliation{%
  \institution{Chinese Academy of Sciences}
  \city{Beijing}
  \country{China}}
\email{zhuangfuzhen@ict.ac.cn}

\author{Jiajie Xu}
\affiliation{%
  \institution{Soochow University}
  \city{Suzhou}
  \country{China}}
\email{xujj@suda.edu.cn}

\author{Xiaofang Zhou}
\affiliation{%
  \institution{The University of Queensland}
  \city{Queensland}
  \country{Australia}}
\email{zxf@itee.uq.edu.au}

\author{Hui	Xiong}
\affiliation{%
  \institution{Rutgers University New Jersey}
  \city{New Jersey}
  \country{USA}}
\email{hxiong@rutgers.edu}


 





%
\begin{abstract}
When purchasing appearance-first products, e.g., clothes, product appearance aesthetics plays an important role in the decision process. 
Moreover, user's aesthetic preference, which can be regarded as a personality trait and a basic requirement, is domain independent and could be used as a bridge between domains for knowledge transfer. 
However, existing work has rarely considered the aesthetic information in product photos for cross-domain recommendation.
To this end, in this paper, we propose a new deep Aesthetic preference Cross-Domain Network (ACDN), in which parameters characterizing personal aesthetic preferences are shared across networks to transfer knowledge between domains. Specifically, we first leverage an aesthetic network to extract relevant features. Then, we integrate the aesthetic features into a cross-domain network to transfer users' domain independent aesthetic preferences. Moreover, network cross-connections are introduced to enable dual knowledge transfer across domains. Finally, the experimental results on real-world data show that our proposed ACDN outperforms other benchmark methods in terms of recommendation accuracy. The results also show that users' aesthetic preferences are effective in alleviating the data sparsity issue on the cross-domain recommendation.

\end{abstract}

%
%


%

%

%
\maketitle
\input{introduction}
\input{relatedwork}
\input{preliminary}

\input{Model}
\input{experiment.tex}

\input{conclusion}
\input{reference.bbl}

\bibliographystyle{ACM-Reference-Format}
\end{document}

%% file: introduction.tex
\section{Introduction}
Recommendation systems have attracted a great amount of interests in recent years. They are utilized to handle the information overload problem and help people make right decisions according to their historical behaviors. 
When shopping online, we usually look through product images before making the decision, especially products that are important in appearance, e.g., clothes, shoes. Product images provide abundant visual information, including design, color schemes, decorative patterns, texture, and so on. We can even estimate the quality and the authenticity of a product from its images. As such, visual information plays an important role in improving the performance of recommendation with appearance priority.

Researchers have started to use image data for recommendation with various image features, such as features extracted by convolutional neural networks (CNN features), the scale-invariant feature transform algorithm (SIFT features), and color histograms \cite{Zhao:2016,Zhao:2017,He:2016}. 
These image features contain semantic information to distinguish items and have been proved effective in recommendation tasks. However, one important visual factor, aesthetics, has rarely been considered in previous visual content enhanced recommendation systems. When purchasing appearance-first products, what consumers concern is not only "What is the product?", but also "Does the product look good?" and "Does the product match the aesthetic preference?". 
Unfortunately, the image features, e.g., CNN features and SIFT features, do not encode aesthetic information by nature. Thus, to provide a high-quality recommendation, comprehensive and high-level aesthetic features are greatly desired.

Image aesthetics assessment, which requires an in-depth understanding of photographic attributes and semantics in an image, has a variety of applications, such as image search, photo ranking, and personal album curation. 
To characterize the complex and personal aesthetic perception, increasing research interests can be observed. For example, deep aesthetic networks have been developed to imitate human  aesthetic perception and achieve the ability to represent image content from low-level features to high-level features \cite{BDN2016,Lu:2014,MaCVPR2017}. 
It is easy to understand that image aesthetics is a highly subjective task as individual user has very diversified aesthetic preferences. For instance, 
some people like the simple black and white appearance products, while some like colorful, flowery and punk style products, and some others like the outdoor wild wind style.
Hence, if you know more about your consumer's aesthetic preferences, you can recommend the products more convincingly according to consumer's taste. 
However, few efforts have been found considering the aesthetic preferences for recommendation except Yu et al. \cite{yu2018aesthetic} introduce aesthetic information into clothing recommendation systems. It demonstrates that incorporating aesthetic features can improve the recommendation performance significantly, since aesthetic features and CNN features complement each other. However, it does not consider aesthetic features for cross-domain recommendation.
\begin{figure}
\setlength{\belowcaptionskip}{-0.5cm}   
    \centering
    \includegraphics[width=3.3in]{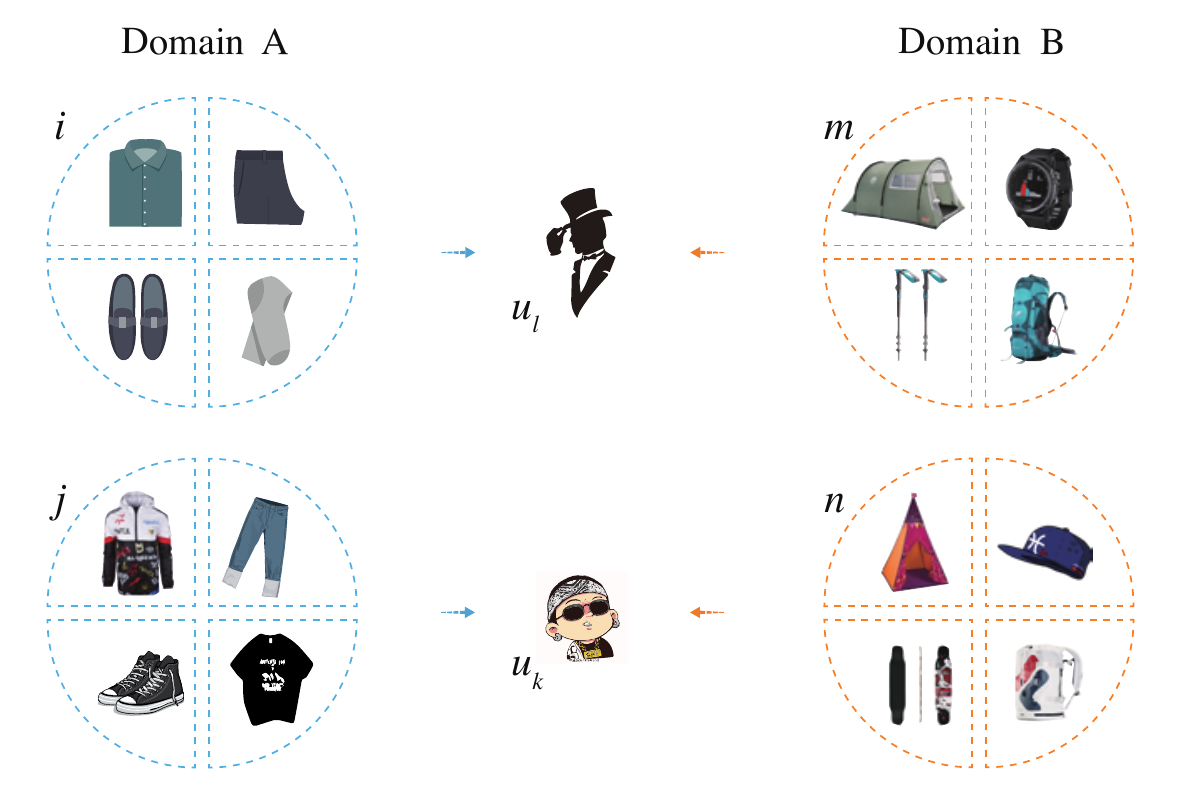}
    \caption{A user's aesthetic behaviors in different areas have consistency.}
    \label{fig:intro}
\end{figure}

Generally, users are active in many E-commerce websites and have a large number of behavioral data in different domains. And the aesthetic preference varies significantly from user to user. However, a user's aesthetic behaviors in different areas could be consistent.  
For example, as shown in Figure \ref{fig:intro},  if a user $u_l$ likes the simple black and white style, she/he will prefer item $i$ in domain $A$ and item $m$ in domain $B$. If a user $u_k$ likes bells and whistles, a hip hop style, she/he will prefer item $j$ in domain $A$ and item $n$ in domain $B$.
Based on the above observation, we can see the aesthetic behavioral data of domain $A$ may help model the aesthetic preferences in domain $B$.
This consistency of aesthetic behavior is helpful for cross-domain recommendation, especially when one domain suffers from the data sparsity issue.

To capture aesthetic preferences and to transfer knowledge among different domains, we propose a new deep Aesthetic preference Cross-Domain Network, termed as ACDN, in which parameters characterizing the personal aesthetic preferences are shared across different domains to achieve a significant improvement for recommendation. 
Specifically, we first leverage an aesthetic network to extract relevant features. 
We utilize a deep aesthetic network (i.e., ILGNet \cite{jin2018ilgnet}) to extract the holistic features to represent the aesthetic elements of a product photo (for example, the aesthetic elements can be color, structure, proportion, style, etc.).
Then, we incorporate the aesthetic features into a deep cross-domain recommendation network. 
Moreover, dual knowledge transfer is achieved by using dual cross transfer unit and joint loss function, which can enable them benefit from each other. 
Finally, we conduct extensive experiments to evaluate the effectiveness of the proposed model ACDN on two real-world Amazon datasets. Our experimental results show that ACDN achieves better performance in terms of the ranking metric, comparing with various baselines. We conduct a thorough analysis to understand how the aesthetic features and transferred knowledge help improve the performance of ACDN.

To the best of our knowledge, ACDN is the first deep model that transfers knowledge from auxiliary domain for recommendation with the aesthetic preference. The main contributions of this work are summarized as follows.
\begin{itemize}
	\item We leverage novel aesthetic features for cross-domain recommendation to capture users' domain independent aesthetic preferences. Moreover, we compare the effectiveness of the aesthetic features with different types of conventional features for cross-domain recommendation to demonstrate the advantage of the aesthetic features.
	\item We propose a new cross-domain recommendation algorithm ACDN for better modeling an individual's propensity from the aesthetic perspective for recommendation, in which the aesthetic preference of each individual is shared for knowledge transfer across different domains.
	\item We conduct extensive experiments on two real-world cross-domain datasets. Our experimental results show the proposed model ACDN outperforms the state-of-the-art methods via comprehensive analysis. Moreover, it can alleviate the data sparsity issue.
\end{itemize}

The remainder of this paper is organized as follows. 
Section \ref{RW} briefly introduces the related work. 
Section \ref{Preliminary} provides the Notations and problem definition. 
In Section \ref{model}, we introduce our proposed ACDN model in detail. 
Our experimental results with analysis are shown in Section \ref{exp}. 
Finally, we conclude this paper in Section \ref{conclusion}.

%% file: relatedwork.tex
\section{Related Work}\label{RW}

\subsection{Collaborative Filtering}
Recommender system is usually seen as predicting users' preferences on unobserved items based on their past history interactions. Collaborative filtering (CF) is an early popular and widely used recommendation method based on matching users with similar tastes or interests \cite{herlocker1999algorithmic}. 
One representative technology for CF is Matrix Factorization (MF), which learns latent factors of users 
and items from a user-item rating matrix \cite{mnih2008probabilistic,koren2009matrix}. Latent factor models extract feature vectors for users and items mainly based on MF. Factorization Machine (FM) can mimic MF with the flexibility of feature engineering \cite{rendle2012factorization}. Moreover, with the 
revival of neural networks, neural CF methods are proposed to learn the underlying complex user-item 
interactions with a highly nonlinear function, such as Wide $\&$ Deep \cite{cheng2016wide} and 
NCF \cite{he2017neural}. However, these CF-based methods based on the sole rating matrix are faced with data sparse and the cold-start problem.

Items are related with content information in general, such as unstructured text and visual features. A famous saying that \textit{A picture is worth a thousand words} suggests that image contains rich information, which is an effective strategy to solve the above problems for recommender system.
For instance, He et al. \cite{He:2016} proposed a scalable factorization model to incorporate visual features from product images into predictors of people's opinions. Zhao et al. \cite{Zhao:2017} proposed a visual-enhanced probabilistic matrix factorization model for tour recommendation, which integrates visual features into the collaborative filtering model. Recently, Yu et al. \cite{yu2018aesthetic} proposed
a coupled matrix and tensor factorization model for aesthetic-based clothing recommendation in which CNNs are used to learn the image features and aesthetic features. Different from our work, the above methods only focus on single domain recommendation.

\subsection{Cross-domain Recommendation}
Cross-Domain Recommendation (CDR) \cite{cremonesi2011cross} is another effective technique for alleviating data sparse issues by leveraging the rating information from other domains to enhance the performance on the target domain \cite{Zhang2012Multi}. Existing CDR methods can be divided into two groups, i.e., content-based 
and transfer-based. 
Berkovsky et al. \cite{berkovsky2007cross} proposed a content-based CDR approach targeting the data sparsity problem by importing and aggregating vectors of users' ratings operating in different application domains. Later on, Winoto et al. \cite{winoto2008if} 
uncovers the association between user preferences on related items across domains. 
Transfer-based approaches mainly employ machine learning techniques (e.g., transfer learning and neural networks) to transfer knowledge across domains. Li et al. \cite{Li2009Can} proposed a codebook method, 
which transfers user-item rating patterns from an auxiliary task in other domains to a sparse rating matrix in a target domain. Man et al. \cite{ijcai2017-343} proposed an embedding and mapping framework (EMCDR), 
which uses a multi-layer perceptron to learn the nonlinear mapping function between a source domain and a target domain. In terms of neural network, Misra et al. \cite{misra2016cross} proposed a convolutional network with cross-stitch units to learn an optimal combination of shared and task-specific representation using multi-task 
learning, and hence enable the knowledge transfer between two domains.
However, these methods treat knowledge transfer as a global process with shared global parameters and do not match source items with the specific target item given a user. 
Different from the above works, we introduce novel aesthetic features for cross-domain recommendation to capture users' domain independent aesthetic preference and propose a new deep aesthetic preference cross-domain network for better modeling an individual's propensity from the aesthetic perspective for recommendation.

%% file: preliminary.tex
\section{Notations and Problem Definition}\label{Preliminary}
In this section, we will introduce related notations and our problem settings.
Given a target domain $\mathcal{T}$ and a source domain $\mathcal{S}$, where
users $\mathcal{U}$ (its size
$m$ = |$\mathcal{U}$|) are shared, we want to transfer knowledge
across domains. We denote the set of items in source domain $\mathcal{S}$ as $I_{S}$ and the size of items in source domain is $n_{S}$ = $|I_{S}|$.
Similarly, we denote the set of items in target domain $\mathcal{T}$ as $I_{T}$ and 
its size is $n_{T}$ = $|I_{T}|$.
We use $u$ to index a user, $i$ to index a target item and $j$ to index a source item. Then, matrix $\boldsymbol{R}_T \in \mathbb{R}^{m \times n_T}$ is used to represent the user-item interaction matrix in the target domain, and the entry
$r_{ui} \in \{0,1\}$ is $1$ if the user $u$ has purchased the item $i$ and $0$ otherwise. Similarly for the source domain,
matrix $\boldsymbol{R}_S \in \mathbb{R}^{m \times n_S}$ is used to describe user-item interactions, the
entry $r_{uj} \in \{0,1\}$ is $1$ if user $u$ has an interaction with item $j$ and $0$ otherwise.
Here each domain can be treated as a problem of collaborative filtering for implicit feedback ~\cite{hu2008collaborative,pan2008one}.

For the task of item recommendation, our goal is to recommend a ranked list of items for each user based
on his/her history records, i.e., top-$N$ recommendation. We aim to improve the recommendation performance
in the target domain with the help of the user-item interaction information and user's aesthetic preference from the source domain. The items are ranked by their predicted scores:
\begin{equation}
   \hat{r}_{ui} = f(u, i|\Theta),
\end{equation}
where $f$ is an interaction function and $\Theta$ are model parameters. For matrix factorization
techniques, the match function is the fixed dot product:
\begin{equation}
   \hat{r}_{ui} = \boldsymbol{P}_{u}^T \boldsymbol{Q}_i,
\end{equation}
and parameters $\Theta = \{\boldsymbol{P}, \boldsymbol{Q}\}$ are latent vectors of users and items,
where $\boldsymbol{P} \in \mathbb{R}^{m \times d}$, $\boldsymbol{Q} \in \mathbb{R}^{n \times d}$ and
$d$ is the dimension size. For neural CF approaches, neural networks are used to a parameterized function $f$
and learn it from interactions:
\begin{equation}\label{FFNN}
   f(\boldsymbol{x}_{ui}|\boldsymbol{P},\boldsymbol{Q},\theta_f) = \phi_o(\phi_L(...(\phi_1(\boldsymbol{x}_{ui})))),
\end{equation}
where the input $\boldsymbol{x}_{ui} = [\boldsymbol{P}^T \mathbb{X}_u, \boldsymbol{Q}^T \mathbb{X}_i]$ is
merged from projections of the user and the item, and the projections are based on their one-hot encodings
$\mathbb{X}_u\in \{0, 1\}^m, \mathbb{X}_i \in \{0, 1\}^n$ and embedding matrices $\boldsymbol{P} \in \mathbb{R}^{m \times d}$,
$\boldsymbol{Q} \in \mathbb{R}^{n \times d}$. The output and the hidden layers are computed by $\phi_o$ and $\phi_l$ ($l \in [1,L]$)
in a multi-layer feedforward neural network (FFNN), and the connection weight matrices and biases are denoted
by $\theta_f$.

In our aesthetic preference cross-domain recommendation network, each domain is modeled by a neural
network, and these networks are jointly learned to improve the performance through mutual knowledge transfer.

%% file: Model.tex
\section{The Proposed MODEl}\label{model}
\subsection{Model Overview}
In this subsection, we briefly describe the proposed Aesthetic preference Cross-Domain Network model (ACDN), in which parameters characterizing the personal aesthetic preferences are shared across different domains to achieve a significant improvement for cross-domain recommendation.
\begin{figure*}
\centering
	\includegraphics[width=7.0in]{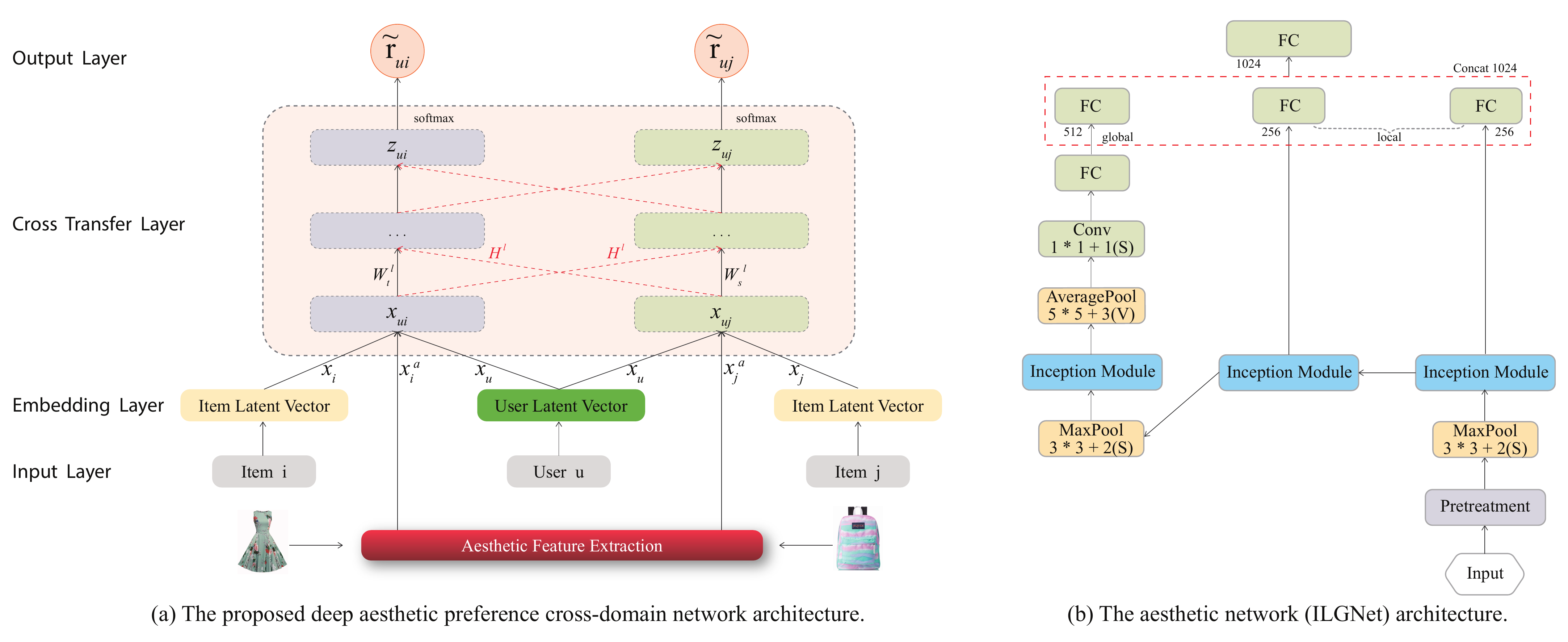}
	\caption{The left figure is the proposed deep aesthetic preference cross-domain model architecture and the right figure is the aesthetic network (ILGNet) architecture.}
	\label{model_fig}
\end{figure*}


As is shown in Figure \ref{model_fig}(a), we adopt FFNN as the base network for each domain to parameterize the interaction function. The base network is similar to the Deep Model in \cite{cheng2016wide,covington2016deep} and the MLP model in [12]. The proposed ACDN model
processes the information flow from the input
to the output with following four modules: Aesthetic Feature Extraction, Embedding Layer, Cross Transfer Layer and Output Layer.
On the bottom of the figure is aesthetic feature extraction. For each item $i$ in the target domain and item $j$ in the source domain, we utilize the pre-trained deep aesthetic network to extract the aesthetic features from a corresponding image in advance.
In the embedding layer, 
we embed the sparse one-hot encoding representation into
a dense vector. The obtained user (item) embedding can
be seen as the latent vector for user (item) in the context
of the latent factor model. Then, the user embedding, item embedding, and aesthetic features are concatenated.
Above the embedding layer is the cross transfer layer, which 
can enable dual knowledge transfer across domains from one base network to another and vice versa. The core idea of the cross transfer unit is to adopt a relationship/transfer matrix rather than a scalar weight to transfer knowledge. We enforce a sparse structure ($l_1$-norm regularization) on the relationship/transfer matrix to control knowledge transfer so that the cross transfer layer can adaptively
transfer selective and useful information.
The final output layer is used to predict the score $\hat{r}_{ui}$
for the given user-item pair based on the representation $z_{ui}$ from the last layer of the multi-hop module.
In the following subsections, we will introduce our model in detail. 
\subsection{Aesthetic Feature Extraction}
We utilize the pre-trained deep aesthetic neural network ILGNet \cite{jin2018ilgnet} to extract aesthetic features from item images. 
ILGNet (I : Inception, L : Local, G : Global) is a novel deep convolutional neural network , which introduces the inception module into image aesthetics classification and can extract aesthetic features from low level to high level.
As is shown in Figure \ref{model_fig}(b),
this network connects the layer of local features to the layer of global features to form a concat layer of 1024 dimension, which are binary patterns. 
Specifically, 
the first and the second inception layers are considered
to extract local image features and the last inception layer is
considered to extract global image features after two max
pooling and one average pooling. Then, we connect the output
of the first two inception layers (256 dimension for each) and
the last inception layer (512 dimension) to form a 1024 dimension
concat layer as the holistic aesthetic feature.

In our work, for each item $i$ in the target domain, we utilize the pre-trained ILGNet\footnote{https://github.com/Kin-Lau/ILGnet} to extract
its aesthetic features $ \boldsymbol{x}_i^a \in \mathbb{R}^{1 \times 1024}$ from the corresponding image in advance.
Similarly, for each item $j$ in the source domain, we obtain its aesthetic feature $\boldsymbol{x}_j^a \in \mathbb{R}^{1 \times 1024}$.
With the aesthetic features of items, we can capture users' aesthetic preference across domains and improve the target domain recommendation performance.

\subsection{Embedding Layer}
To represent the input, we encode user-item interaction indices by one-hot encoding. For user $u$, item $i$ from the target domain and item $j$ from the source domain, we map them into one-hot encoding $\mathbb{X}_u \in \{0, 1\}^m$, $\mathbb{X}_i \in \{0, 1\}^{n_T}$ and  $\mathbb{X}_j \in \{0, 1\}^{n_S}$, where only
the element corresponding to index is $1$ and others are $0$. 
Then, we embed one-hot encodings 
into continuous representation $\boldsymbol{x}_u = \boldsymbol{P}^T\mathbb{X}_u$, $\boldsymbol{x}_i = \boldsymbol{Q}_t^T\mathbb{X}_i$ and $\boldsymbol{x}_j = \boldsymbol{Q}_s^T\mathbb{X}_j$ by
embedding matrices $\boldsymbol{P}$, $\boldsymbol{Q_t}$ and $\boldsymbol{Q_s}$, respectively. 
Finally, we concatenate $\boldsymbol{x}_{ui} = [\boldsymbol{x}_u,\boldsymbol{x}_i,\boldsymbol{x}_i^a]$, $\boldsymbol{x}_{uj} = [\boldsymbol{x}_u,\boldsymbol{x}_j,\boldsymbol{x}_j^a]$ to be the 
input of following building blocks.

\subsection{Cross Transfer Layer}
In this subsection, we will introduce the cross transfer layer for knowledge transfer in detail. Different from CSN \cite{misra2016cross}, the core idea of the cross transfer unit is to adopt a relationship/transfer matrix rather than a scalar weight to transfer knowledge. The target domain can receive information from the source domain and vice versa.

As is shown in Figure \ref{model_fig}(a), we add cross transfer units to the entire FFNN.
Denote $\boldsymbol{W}_t^l$ as the weight connecting from the $l$-th layer to the ($l+1$)-th layer and $b_t^l$ as the bias in target domain. Similarly, there are $\boldsymbol{W}_s^l$ and $b_s^l$ in the source domain.
Denote $\boldsymbol{H}^l$ as the relationship matrix from the $l$-th layer to the $(l+1)$-th layer. 
The two base networks can be coupled by cross transfer unit:
\begin{equation}
   \alpha_t^{l+1} = \sigma(\boldsymbol{W}_t^l\alpha_t^l + \boldsymbol{b}_t^l + \boldsymbol{H}^l\alpha_s^l),
\end{equation}
\begin{equation}
    \alpha_s^{l+1} = \sigma(\boldsymbol{W}_s^l\alpha_s^l + \boldsymbol{b}_s^l + \boldsymbol{H}^l\alpha_t^l),
\end{equation}
where $\sigma$ is the activation function and we use ReLU \cite{nair2010rectified} here.
In the target domain, we can observe that the representations of the $(l+1)$-th layer $\alpha_t^{l+1}$ receives two
information flows: one is from the transform gate controlled by a weight matrix 
$\boldsymbol{W}_t^l$ and another is from transfer gate controlled by 
$\boldsymbol{H}^l$
(similarly for the $\alpha_s^{l+1}$ in the source domain). 
This way of knowledge transfer happens in two directions, 
from the source domain to the target domain and from the target domain to the source domain, which can enable dual knowledge transfer across domains and let them benefit from each other.
Similar to CSN \cite{misra2016cross},
we take the same relationship/transfer matrix $\boldsymbol{H}^l$ for 
both directions to reduce model parameters and make the model compact. 
Actually, it does not improve the performance of recommendation by taking different transfer matrices for two directions.

Obviously, the relationship/transfer matrix $\boldsymbol{H}^l$ is very crucial to our model. We assume that not all representations from another domain are useful and we expect that the representations receiving 
from other domains are selective and useful. 
This corresponds to enforcing a sparse prior on the 
structure and can be achieved by penalizing the relationship/transfer matrix $\boldsymbol{H}^l$ via regularization. 
We take the widely used sparsity-induced regularization: least absolute shrinkage and selection 
operator \cite{tibshirani1996regression}. 
We enforce the $l_1$-norm regularization on the relationship/transfer matrix $\boldsymbol{H}^l$ to induce sparsity:
\begin{equation}\label{H}
   \Omega(\boldsymbol{H}^l) = \lambda\sum_{i=1}^r \sum_{j=1}^q |h_{i,j}|,
\end{equation}
where $h_{ij}$ is the entry ($i, j$)
of $\boldsymbol{H}^l$, hyper-parameter $\lambda$ controls the degree of sparsity and
$r \times q$ is the size of matrix 
$\boldsymbol{H}^l$. It means that $\boldsymbol{H}^l$ linearly transforms representations
$\alpha_s^l \in \mathbb{R}^q $ in the source domain and the result is as part of the input to the next
layer $\alpha_t^{l+1} \in \mathbb{R}^r$ in the target domain.

\subsection{Model Learning}
According to the task of item recommendation and the nature of the implicit feedback, 
we adopt cross-entropy as our loss function for model
optimization. The objective function to be minimized in the model
optimization is defined as follows:
\begin{equation}\label{base loss}
  \mathcal{L}_0 = -\sum_{(u,i)\in \mathbf{R}^+ \cup \mathbf{R}^-} r_{ui} log \hat{r}_{ui} +(1-r_{ui})log(1-\hat{r}_{ui}),
\end{equation}
where $\mathbf{R}^+$ and $\mathbf{R}^-$ are the observed interaction matrix and randomly sampled negative examples \cite{pan2008one}, respectively.
This objective function has probabilistic interpretation and is the negative logarithm likelihood of the following likelihood
function:
\begin{equation}
  \mathcal{L}(\Theta|\mathbf{R}^+ \cup \mathbf{R}^-) = \prod_{(u,i)\in \mathbf{R}^+}\hat{r}_{ui}\prod_{(u,i)\in\mathbf{R}^-}(1-\hat{r}_{ui}),
\end{equation}
where $\Theta$ are model parameters.

we add \textit{joint loss function} to our proposed model, which can be 
trained efficiently by back-propagation. 
Instantiating the base loss $\mathcal{L}_0$ described in Eq.\ref{base loss} by the loss of the target domain ($\mathcal{L}_T$) and 
loss of the source domain ($\mathcal{L}_s$), the objective function of our proposed model is their joint losses:
\begin{equation}
  \mathcal{L}(\Theta) = \mathcal{L}_t(\Theta_t) + \mathcal{L}_s(\Theta_s),
\end{equation}
where the model parameters $\Theta = \Theta_t \cup \Theta_s$.
This objective function can be optimized by stochastic gradient descent (SGD):
\begin{equation}
  \Theta' \leftarrow \Theta - \eta\frac{\partial \mathcal{L}(\Theta)}{\partial\Theta},
\end{equation}
where $\eta$ is the learning rate. 

\subsection{Complexity Analysis}
The model parameters include $\{\boldsymbol{P}, (\boldsymbol{H}^l)_1^L \}$,
$\{\boldsymbol{Q}_t,(\boldsymbol{W}_t^l,\boldsymbol{b}_t^l)_{l=1}^L\}$ and $\{\boldsymbol{Q}_s,(\boldsymbol{W}_s^l,\boldsymbol{b}_s^l)_{l=1}^L\}$,
where user embedding $\boldsymbol{P}$, item embedding $\boldsymbol{Q}_t$ and $\boldsymbol{Q}_s$ contain numbers of parameters because
they depend on the input size of the user latent vector, the item latent vector and the aesthetic features. Usually, the number of neurons in a hidden layer is 
about one hundred. Thus, the size of the weight matrix and the cross transfer matrix is 
hundreds by hundreds. All in all, the size of model parameters is close to the size of typical latent factor models \cite{koren2009matrix} and is linear with the input size.
During training process, we update the target network and the source network by the data of the corresponding domain. The learning strategy is similar to CSN \cite{misra2016cross} and the total cost of learning each base network is approximately equal to that of running a typical neural 
CF approach \cite{he2017neural}. Totally, the whole network can be trained efficiently by back-propagation with mini-batch stochastic 
optimization.
\begin{table*}[ht]
	\caption{Performance Comparison of different methods on two datasets. 
	The best performance is highlighted in boldface.}
	\centering
  \small
  \label{Result_table}
  \resizebox{\textwidth}{25mm}{
  \begin{tabular}{c|lllllllll|lll|lll|lll}
  \toprule[1pt]
  \textbf{Dataset}                 & \multicolumn{9}{c|}{\textbf{Clothing \& Home Improvement (Dataset 1)}}                                                                                                                                                                                                                                  & \multicolumn{9}{c}{\textbf{Outdoor and Sports \& Clothing (Dataset 2)}}                                                                                                                                                                                                     \\ \hline
  \multirow{2}{*}{\textbf{Method}} & \multicolumn{3}{c|}{\textit{TopN = 5}}                                                & \multicolumn{3}{c|}{\textit{TopN = 10}}                                               & \multicolumn{3}{c|}{\textit{TopN = 20}}                                               & \multicolumn{3}{c|}{\textit{TopN = 5}}                                       & \multicolumn{3}{c|}{\textit{TopN = 10}}                                      & \multicolumn{3}{c}{\textit{TopN = 20}}                                     \\
                                    & \multicolumn{1}{c}{HR}     & \multicolumn{1}{c}{NDCG}   & \multicolumn{1}{c|}{MRR}    & \multicolumn{1}{c}{HR}     & \multicolumn{1}{c}{NDCG}   & \multicolumn{1}{c|}{MRR}    & \multicolumn{1}{c}{HR}     & \multicolumn{1}{c}{NDCG}   & \multicolumn{1}{c|}{MRR}    & \multicolumn{1}{c}{HR} & \multicolumn{1}{c}{NDCG} & \multicolumn{1}{c|}{MRR} & \multicolumn{1}{c}{HR} & \multicolumn{1}{c}{NDCG} & \multicolumn{1}{c|}{MRR} & \multicolumn{1}{c}{HR} & \multicolumn{1}{c}{NDCG} & \multicolumn{1}{c}{MRR} \\ \hline \hline
  BPRMF                            & 0.0902                     & 0.0753                     & \multicolumn{1}{l|}{0.0650} & 0.1730                     & 0.0941                     & \multicolumn{1}{l|}{0.0704} & 0.2757                     & 0.0939                     & 0.0785                      &0.1105                        &0.0853                          &0.0766                          & 0.1743                       & 0.0975                         & 0.0779                         &0.2848                        &0.1371                          & 0.0892                 \\
  VBPR                             &  0.1027                          &0.0831                            & \multicolumn{1}{l|}{0.0778}       &0.1836                   & 0.1103        & \multicolumn{1}{l|}{0.0831}       
  &0.2903                      &0.1142                   &0.0811                   & 0.1335                       & 0.0970                 &0.0885                &0.1976                        & 0.1105               & 0.0861        &0.3044     &0.1501                        & 0.1023                        \\
  CMF          &0.1201          &0.0903                            & \multicolumn{1}{l|}{0.0812}       
  &0.2014                  &0.1213              & \multicolumn{1}{l|}{0.0947}       & 0.3189                &0.1242                &0.0811                             & 0.1479            &0.1022                 &0.0931              &0.2214         & 0.1233             &0.1005               &0.3237              &0.1601                          &0.1255                         \\
  CDCF     &0.1130                     &0.0863                  & \multicolumn{1}{l|}{0.0794}       &0.1904                 &0.1178                            & \multicolumn{1}{l|}{0.0877}       &0.3054         &0.1183                            &0.0803                             &0.1338                        &0.0928                          & 0.0876                         &0.2103                        &0.1167                          &0.0931                          &0.3155            &0.1566                          & 0.1148                        \\
  MLP                              & \multicolumn{1}{c}{0.1251} & \multicolumn{1}{c}{0.0926} & \multicolumn{1}{c|}{0.0866} & \multicolumn{1}{c}{0.2079} & \multicolumn{1}{c}{0.1225} & \multicolumn{1}{c|}{0.0988} & \multicolumn{1}{c}{0.3266} & \multicolumn{1}{c}{0.1385} & \multicolumn{1}{c|}{0.0871} & \multicolumn{1}{c}{0.1533}     &\multicolumn{1}{c}{0.1047}                   & 0.0958                   & 0.2321                 & 0.1280                   & 0.1021                   & 0.3304                 & 0.1622                   & 0.1295                  \\
  MLP++                            & \multicolumn{1}{c}{0.1292} & \multicolumn{1}{c}{0.0957} & \multicolumn{1}{c|}{0.0974} & \multicolumn{1}{c}{0.2101} & \multicolumn{1}{c}{0.1278} & \multicolumn{1}{c|}{0.1033} & \multicolumn{1}{c}{0.3321} & \multicolumn{1}{c}{0.1379} & \multicolumn{1}{c|}{0.0944} & \multicolumn{1}{c}{0.1590}     &\multicolumn{1}{c}{0.1136}                   & 0.1011                   & 0.2467                 & 0.1339                   & 0.1104                   & 0.3367                 & 0.1734                   & 0.1356                  \\
  CSN                              & \multicolumn{1}{c}{0.1388} & \multicolumn{1}{c}{0.1022} & \multicolumn{1}{c|}{0.0922} & \multicolumn{1}{c}{0.2179} & \multicolumn{1}{c}{0.1335} & \multicolumn{1}{c|}{0.1104} & \multicolumn{1}{c}{0.3465} & \multicolumn{1}{c}{0.1424} & \multicolumn{1}{c|}{0.1027} & \multicolumn{1}{c}{0.1655}     &\multicolumn{1}{c}{0.1243}                   & 0.1033                   & 0.2498                 & 0.1449                   & 0.1170                   & 0.3390                 & 0.1881                   & 0.1408                  \\
  CoNet                            & \multicolumn{1}{c}{0.1437} & \multicolumn{1}{c}{0.1059} & \multicolumn{1}{c|}{0.1014} & \multicolumn{1}{c}{0.2230} & \multicolumn{1}{c}{0.1383} & \multicolumn{1}{c|}{0.1185} & \multicolumn{1}{c}{0.3524} & \multicolumn{1}{c}{0.1513} & \multicolumn{1}{c|}{0.1143} & \multicolumn{1}{c}{0.1739}      & \multicolumn{1}{c}{0.1328}                   & \multicolumn{1}{c|}{0.1124}                   & \multicolumn{1}{c}{0.2539} & \multicolumn{1}{c}{0.1480}  &\multicolumn{1}{c|}{0.1241}           &\multicolumn{1}{c}{0.3437}            & \multicolumn{1}{c}{0.1938}        & \multicolumn{1}{c}{0.1510}                  \\ 
  ACDN                             & \textbf{0.1472}            & \textbf{0.1077}            & \multicolumn{1}{c|}{\textbf{0.1047}}   &\textbf{0.2289}   & \textbf{0.1403}     & \multicolumn{1}{c|}{0.1220} & \textbf{0.3601}            & \textbf{0.1560}            & \textbf{0.1166}             & \textbf{0.1763}                 & \textbf{0.1357}                   & \multicolumn{1}{c|}{\textbf{0.1140}}                   & \textbf{0.2611}                 & \textbf{0.1529}                   & \textbf{0.1254}                  & \textbf{0.3529}                 & \textbf{0.2003}                   & \textbf{0.1543}                  \\ \hline\hline
  Improve                          & \multicolumn{1}{c}{2.4\%}  &\multicolumn{1}{c}{1.69\%}   & \multicolumn{1}{c|}{3.2\%}  & \multicolumn{1}{c}{2.6\%}   & \multicolumn{1}{c}{1.44\%}& \multicolumn{1}{c|}{2.95\%} & \multicolumn{1}{c}{2.18\%} & \multicolumn{1}{c}{3.10\%} & \multicolumn{1}{c|}{2.01\%}  & \multicolumn{1}{c}{1.38\%}   & \multicolumn{1}{c}{2.18\%}                   &\multicolumn{1}{c|}{1.40\%}                   & \multicolumn{1}{c}{2.83\%}        & \multicolumn{1}{c}{3.31\%}        & \multicolumn{1}{c|}{1.04\%}      & \multicolumn{1}{c}{2.68\%}      & \multicolumn{1}{c}{3.35\%}        & \multicolumn{1}{c}{2.18\%}                  \\ 

  \bottomrule[1pt]
  \end{tabular}}
\end{table*}

%% file: experiment.tex
\section{Experiments}\label{exp}
In this section, we first introduce experimental settings. And
then we conduct experiments to answer the following research
questions and validate our technical contributions.\\
\textbf{RQ1}: How does our proposed cross-domain recommender model ACDN perform as compared with state-of-the-art recommendation methods, including single-domain and cross-domain, visual enhanced methods, and deep/shadow methods?\\
\textbf{RQ2}: What are the advantages of the aesthetic features for cross-domain recommendation, compared with other conventional features, such as color histograms and CNN features?\\
\textbf{RQ3}: How do the hyper-parameters affect the performance of the proposed model?

\subsection{Experimental Setup}
\textbf{Dataset.} 
We study the effectiveness of our proposed approach on a real-world public dataset
\textit{Amazon}\footnote{http://jmcauley.ucsd.edu/data/amazon/links.html} with different kinds of domains.
It contains product reviews and metadata from Amazon, including 142.8 million reviews spanning May 1996 - July 2014, and has been used to evaluate the performance of various approaches. Here we use three domains: \textit{Home Improvement, Clothing, Outdoor and Sports,}
and conduct experiments on two datasets with following combinations. The statistics of the two datasets are summarized in Table \ref{dataset}.

\textbf{Clothing \& Home Improvement (Dataset 1) }: \textit{Source domain = Clothing, Target domain = Home Improvement.} The number of the sharing users is 8,673,
and there are 18,442 items, 56,183 interactions and 21,317 items 60,942 interactions in the target domain and the source domain, respectively. Similar to ~\cite{hu2018conet}, we remove users and items with fewer than 5
purchase records. The density of the two domains is 0.035\% and 0.032\% respectively.

\textbf{Outdoor and Sports \& Clothing (Dataset 2)}: \textit{Source domain = Outdoor and Sports, Target domain = Clothing.} The number of the sharing users is 13,164,
and there are 22,465 items, 82,416 interactions and 17,765 items 68,291 interactions in the target domain and the source domain, respectively. Similar to ~\cite{hu2018conet}, we remove users and items with fewer than 5
purchase records. The density of the two domains is 0.029\% and 0.029\% respectively.\\
\begin{table}[]
	\caption{Dataset Description}
	\small
	\label{dataset}
	\begin{tabular}{p{1.05cm}<\centering|p{1.35cm}<\centering|c|c}
	\hline
	\multirow{2}{*}{Dataset}   & \multirow{2}{*}{Statistics} & Source Domain      & Target Domain    \\
										&                             & Clothing           & Home Improvement \\ \hline
	\multirow{4}{*}{Dataset 1} & \#user                      & 8673               & 8673             \\
										& \#item                      & 21317              & 18442            \\
										& \#interactions              & 60942              & 56183            \\
										& \#density                   & 0.032\%            & 0.035\%          \\ \hline\hline
	\multirow{2}{*}{Dataset}   & \multirow{2}{*}{Statistics} & Source Domain      & Target Domain    \\
										&                             & Outdoor and Sports & Clothing         \\ \hline
	\multirow{4}{*}{Dataset 2} & \#user                      & 13164              & 13164            \\
										& \#item                      & 17765              & 22465            \\
										& \#interactions              & 68291              & 82416            \\
										& \#density                   & 0.029\%            & 0.029\%          \\ \hline
	\end{tabular}
	\end{table}
\noindent\textbf{Evaluation Protocol.} For the item recommendation task, the leave-one-out evaluation is widely
used and we follow the protocol in \cite{he2017neural}. It means that we reserve one interaction as
the test item for each user. We determine hyper-parameters by randomly sampling another interaction per
user as the validation set. We follow the common strategy which randomly samples 99 negative
items that are not interacted by the user and then evaluate how well the recommender can rank the test item against these negative ones. Since we aim at $TopN$ item recommendation, the typical evaluation metrics are hit ratio (HR), normalized discounted cumulative gain (NDCG) and mean reciprocal rank (MRR), where the ranked list is cut off at topN = $\{5,10,20\}$. HR intuitively measures whether the reserved test item is present on the 
top-N list, defined as: 
\begin{equation}
    HR = \frac{1}{\mathcal{|U|}}\sum_{u\in\mathcal{U}} \delta(p_u\leq topN),
\end{equation} 
where $p_u$ is the indicator function. NDCG and MRR also account for the rank of the hit position respectively, which are defined as:
\begin{equation}
    NDCG=\frac{1}{|\mathcal{U}|} \sum_{u \in \mathcal{U}} \frac{\log 2}{\log(p_u + 1)}, 
\end{equation}
\begin{equation}
    MRR=\frac{1}{\mathcal{|U|}}\sum_{u\in\mathcal{U}}\frac{1}{p_u}.
\end{equation}
Note that a higher value is better.

\noindent\textbf{Baselines.}
As is shown in Table \ref{baselines}, we compare with various baselines, categorized as single/cross domain and shadow/deep methods.
\begin{figure*}
	\centering
	\subfigure[]{
		\label{size:subfig:a}
		\includegraphics[height=3.5cm,width=4cm]{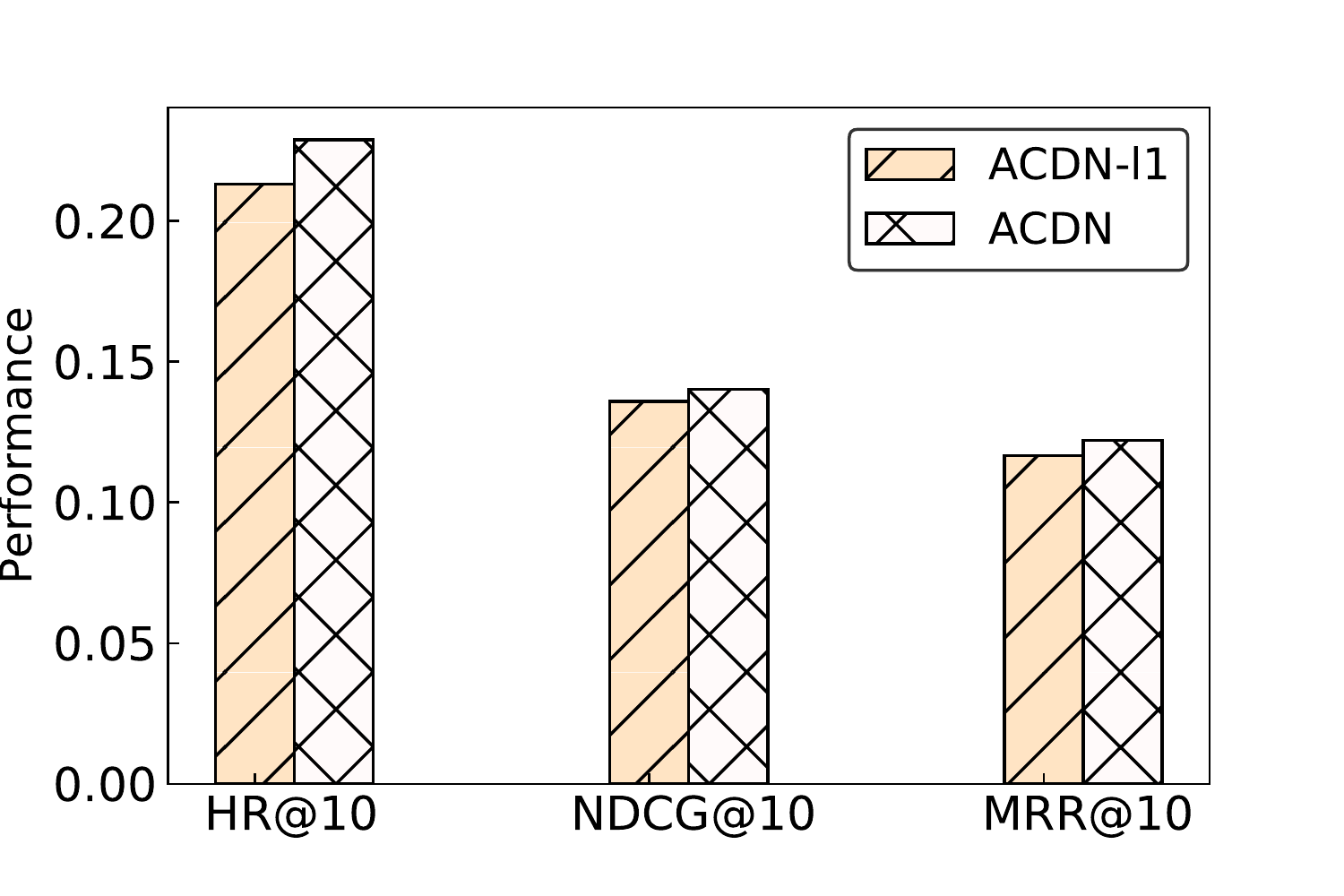}
	}
	\subfigure[]{
		\label{size:subfig:b}
		\includegraphics[height=3.5cm,width=4cm]{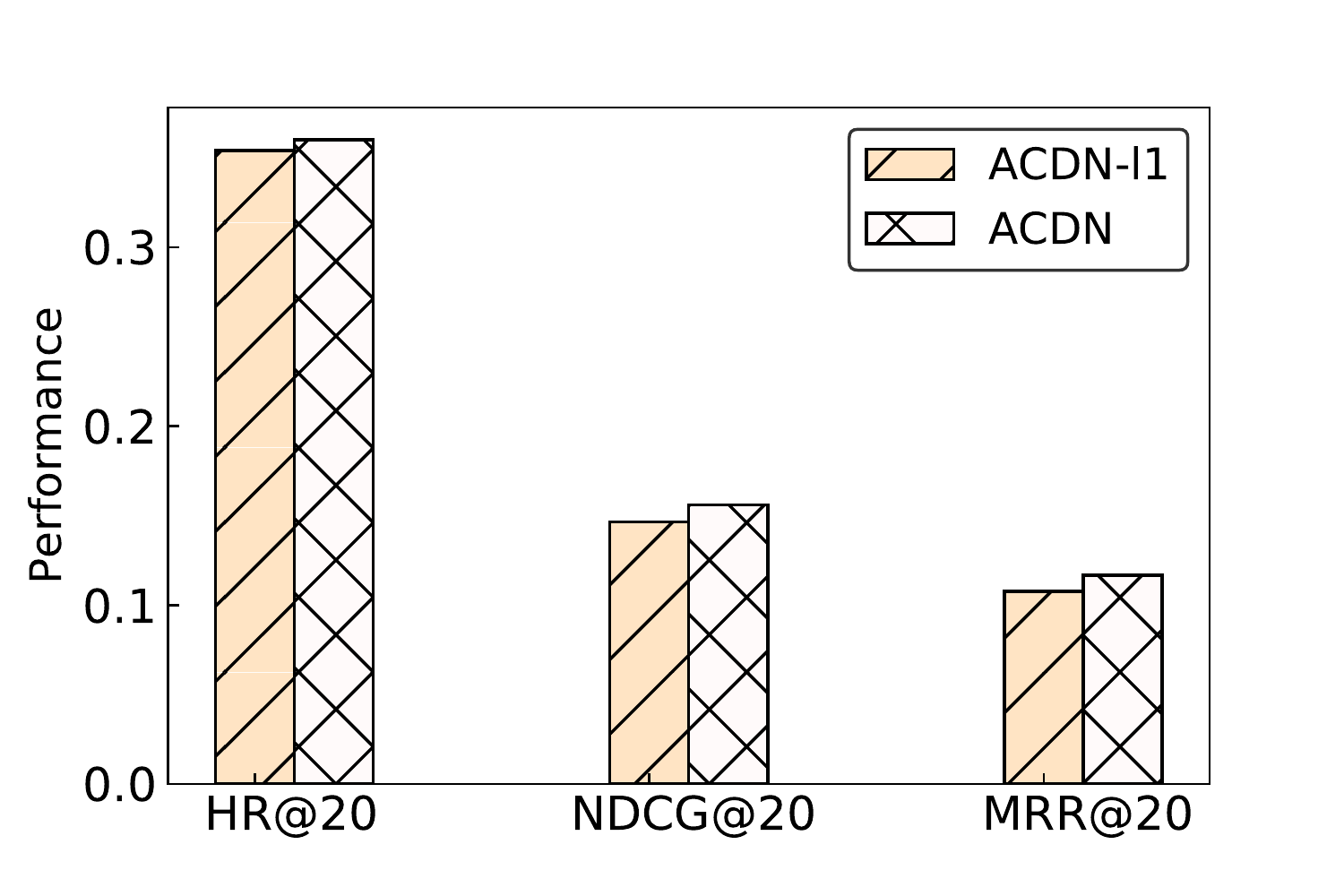}
	}
	\subfigure[]{
		\label{size:subfig:c}
		\includegraphics[height=3.5cm,width=4cm]{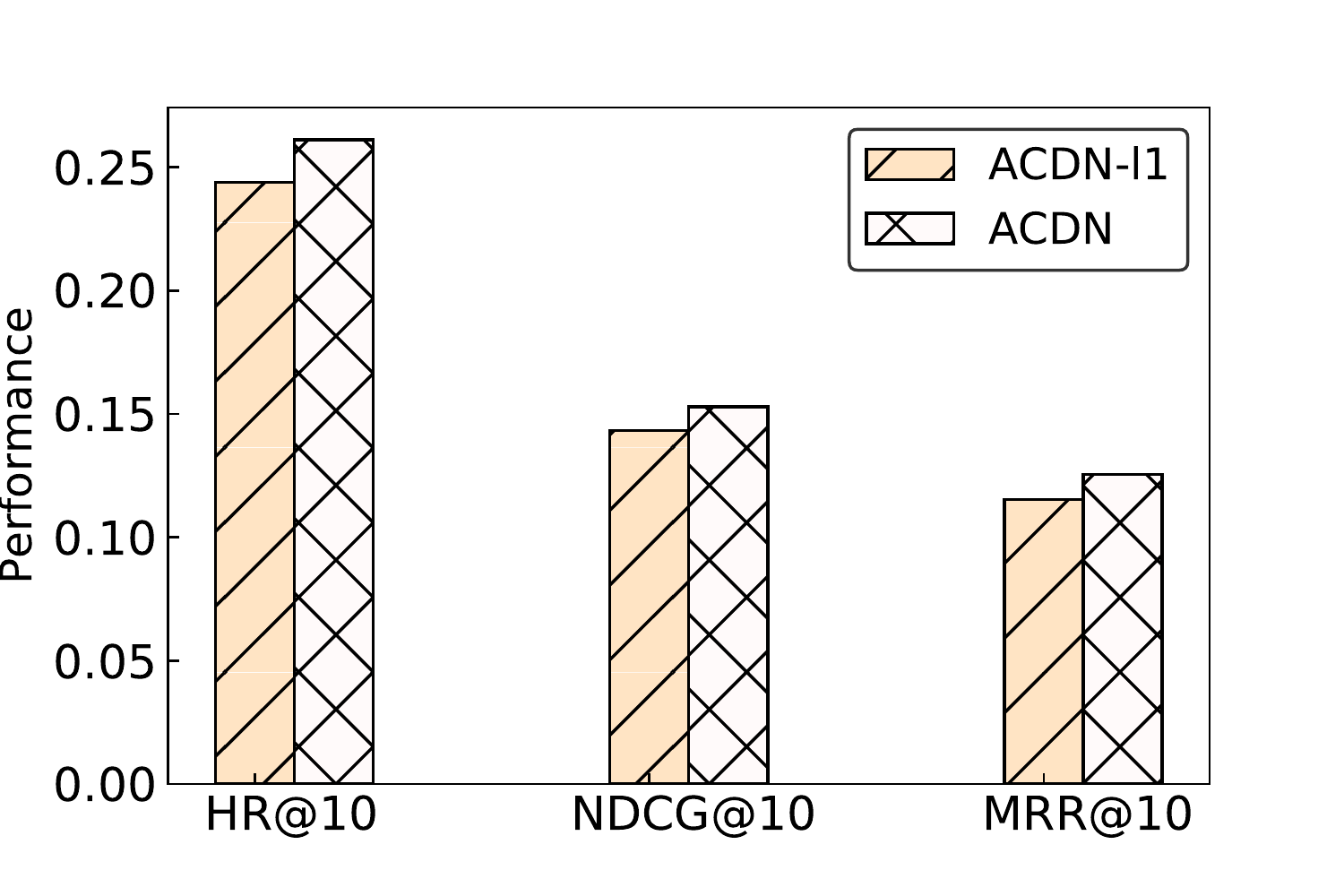}
	}
	\subfigure[]{
		\label{size:subfig:d}
		\includegraphics[height=3.5cm,width=4cm]{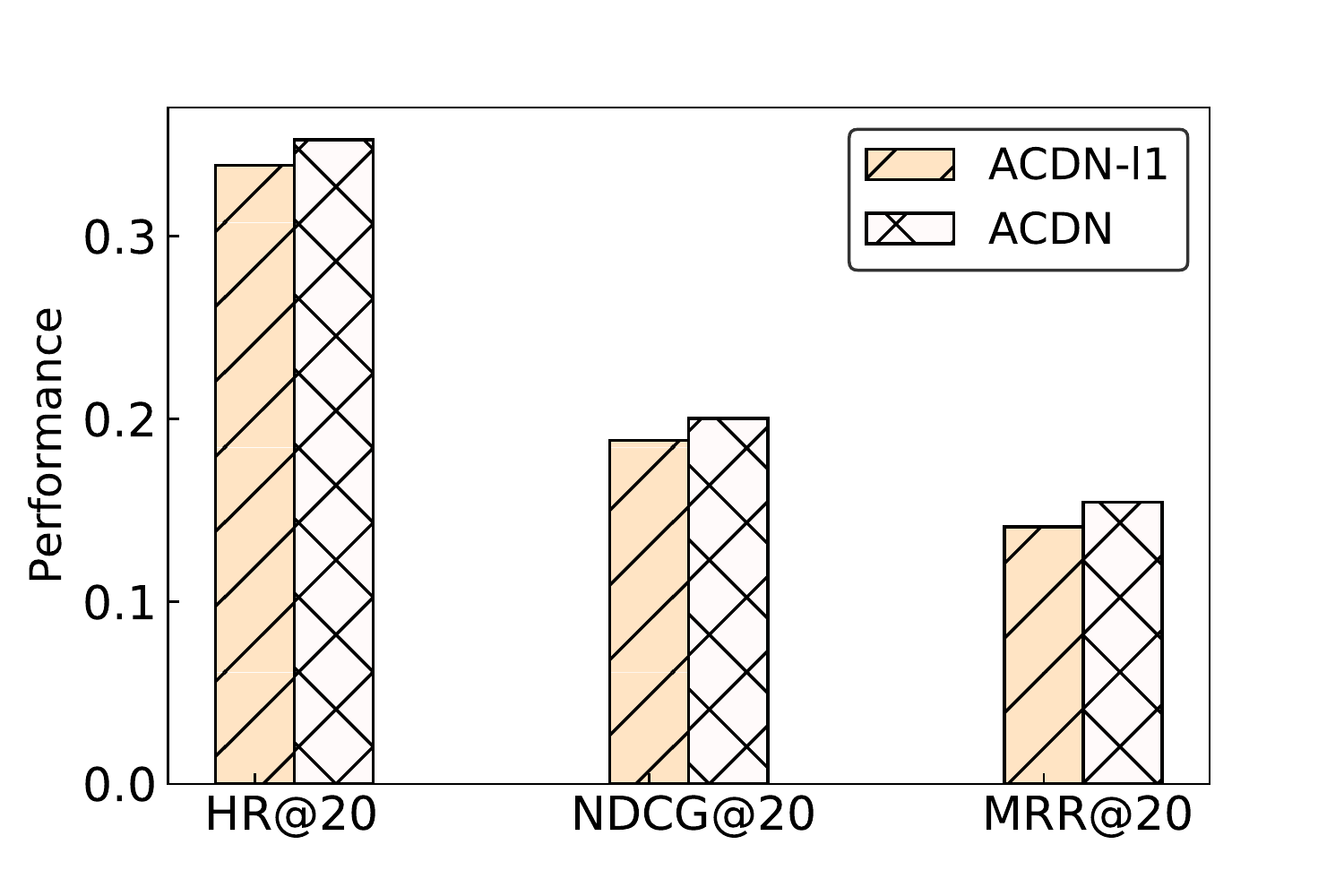}
	}
	\caption{This figure shows the impact of the $l_1$-norm regularization for Sparsity. (a) (b) show the performance on Dataset 1 and (c) (d) show the performance on Dataset 2 in terms of TokN=10, 20.}	
	 \label{sparesize}
\end{figure*}
\begin{itemize}
\item \textbf{BPRMF}: Bayesian personalized ranking \cite{rendle2009bpr} is a typical collaborate filtering approach, which learns the user and item latent factors via matrix 
factorization and pairwise rank loss. 
\item\textbf{MLP}: Multi-layer perception \cite{he2017neural} is a neural collaborate filtering approach, which can learn a user-item interaction function by neural networks. 
\item\textbf{MLP++}: We combine two MLPs by sharing the user embedding matrix. This is a degenerated method that no cross transfer units.
\item\textbf{VBPR}: VBPR \cite{he2016vbpr} is a scalable factorization model to incorporate visual signals into predictors of people's opinions, which can make use of visual
features extracted from product images by pre-trained deep networks. 
\item\textbf{CDCF}: Cross-Domain Collaborate Filtering \cite{loni2014cross} is a cross-domain recommendation method, which is a context-aware approach that applies factorization on the merged domains aligned by the shared users. The auxiliary domain is utilized as a context. 
\item\textbf{CMF}: Collective matrix factorization \cite{singh2008relational} is a multi-relation learning approach, which jointly factorizes matrices of individual 
domains. Here, the relation is user-item interaction. The shared user factors enable knowledge transfer between two domains. 
\item\textbf{CSN}: The cross-stitch network \cite{misra2016cross} is a deep multitask learning model and jointly learns two base networks. It enables knowledge transfer by a linear combination of activation maps from two domains via a shared coefficient. 
\item\textbf{CoNet}: CoNet \cite{hu2018conet} is the latest collaborative cross networks for cross-domain recommendation, which can enable dual knowledge transfer across domains by
introducing cross connections from one base network to another and vice versa and let them benefit from each other.
\end{itemize}
\begin{table}
	\centering
	\small
	\caption{Categories of baselines}
	\label{baselines}
	\begin{small}
		\begin{tabular}{p{2.0cm}<\centering|p{2.3cm}<\centering|c}
			\hline
			Baselines & Shadow method  & Deep method \\\hline\hline
			Single-Domain & BPRMF\cite{rendle2009bpr}VBPR\cite{he2016vbpr} &MLP\cite{he2017neural} \\\hline
			Cross-Domain &CDCF\cite{rendle2012factorization} CMF\cite{singh2008relational} &CoNet\cite{hu2018conet} CSN\cite{misra2016cross} MLP++\\\hline
		\end{tabular}
	\end{small}
\end{table} 

\noindent\textbf{Implementation.}
For BPRMF \cite{rendle2009bpr}, we use LightFM's implementation\footnote{https://github.com/lyst/lightfm}, which is a popular collaborate filtering library. 
For VBPR \cite{he2016vbpr}, we use the open source code\footnote{https://github.com/DevilEEE/VBPR}.
For CDCF \cite{loni2014cross}, we adopt the official libFM implementation\footnote{http://www.libfm.org}. 
For MLP \cite{he2017neural}, we use the code released by its authors\footnote{https://github.com/hexiangnan/neural-collaborative-filtering}. 
For CMF \cite{singh2008relational}, we use a Python version reference to the original Matlab code\footnote{http://www.cs.cmu.edu/~a jit/cmf/}. 
For CSN \cite{misra2016cross}, it requires that the number of neurons in each hidden layer is the same. The configuration can be denoted as [64]$\times$ 4 (means [64, 64 ,64 ,64]).
For CoNet \cite{hu2018conet}, we use the code shared by its author.
Our methods are implemented by Python with TensorFlow and parameters are randomly initialized by Gaussian $\mathcal{N}$(0,0.01). 
We adopt Adam \cite{kingma2014adam} as the optimizer with an initial learning rate 0.001. The ratio of negative sampling is 1 and the size of the mini-batch is 128. As for the design of network structure, we take a tower pattern, having the layer size for each successive higher layer. Specifically, the configuration of hidden layers in each base network is [1152,512,256,128]. 
The size of the first hidden layer(i.e., 1152) is equal to the concatenation of $x_u \in \mathbb{R}^{1\times64}$, $x_i \in \mathbb{R}^{1\times64}$ and $x_i^a \in \mathbb{R}^{1\times1024}$.

\subsection{Performance Comparison (RQ1)}
To demonstrate the recommendation performance of our model ACDN, we compare it with state-of-the-art methods. The experimental results of all methods on two combinations datasets are illustrated in Table \ref{Result_table}, and we have the following observations. 

Firstly, we can find that cross-domain methods (i.e., CMF and CDCF) produce a better performance than single-domain methods (i.e., BPRMF and VBPR) at all settings on both datasets, regardless of shadow methods and deep methods.
This indicates that cross-domain methods benefit from knowledge transfer and is an effective technique for alleviating the data sparsity issue. VBPR outperforms BPRMF, which indicates that visual features extract from item images can indeed enhance the performance of recommendation.

Secondly, we can notice that deep methods perform better than shadow methods in both single-domain and
cross-domain. For example, MLP improves more than 15\% comparing with shadow methods BRPMF and VBPR in all cases in single-domain, and deep cross-domain models (i.e., MLP++, CoNet, and CSN) outperform shadow cross-domain models (i.e., CMF and CDCF) in all cases on two datasets.  This shows the effectiveness of deep neural models with the non-linear combination and more parameters can benefit not only single-domain recommendation but also cross-domain recommendation.

Thirdly, we can observe that our proposed neural model ACDN is better than all baselines on both two datasets at each setting, including the base MLP network, shallow cross-domain models (i.e., CMF and CDCF), deep cross-domain models (i.e., MLP++, CoNet, and CSN). These results demonstrate the effectiveness of the proposed aesthetic features enhanced the cross-domain neural model. 
Comparing MLP++ and MLP, sharing user embedding is
slightly better than the base network due to unilateral knowledge
transfer, which shows the necessity of dual knowledge transfer in a deep way.
CSN is inferior to CoNet on both datasets. The reason is possible that the assumption of CSN is not appropriate: all representations from the auxiliary domain are equally important and are all useful. This motivates us to learn what to transfer adaptively and filter irrelevant information for target domain recommendation
by using a cross transfer matrix rather than a scalar weight.
Also, our model outperforms the state-of-the-art method CoNet since CoNet merely transfers user-item rating information, which demonstrates that aesthetic features can help improve cross-domain recommendation performance, especially in appearance-first products.

In summary, the empirical comparison results demonstrate the superiority of the proposed neural model to transfer aesthetic preference and source domain knowledge for cross-domain recommendation.

\subsection{Necessity of the Aesthetic Features (RQ2)}
\begin{figure}
\setlength{\belowcaptionskip}{-0.6cm}   
   \centering
	\subfigure[Necessity of the Aesthetic Features ]{
      \label{subfig:visual}
		\includegraphics[height=3.5cm,width=4cm]{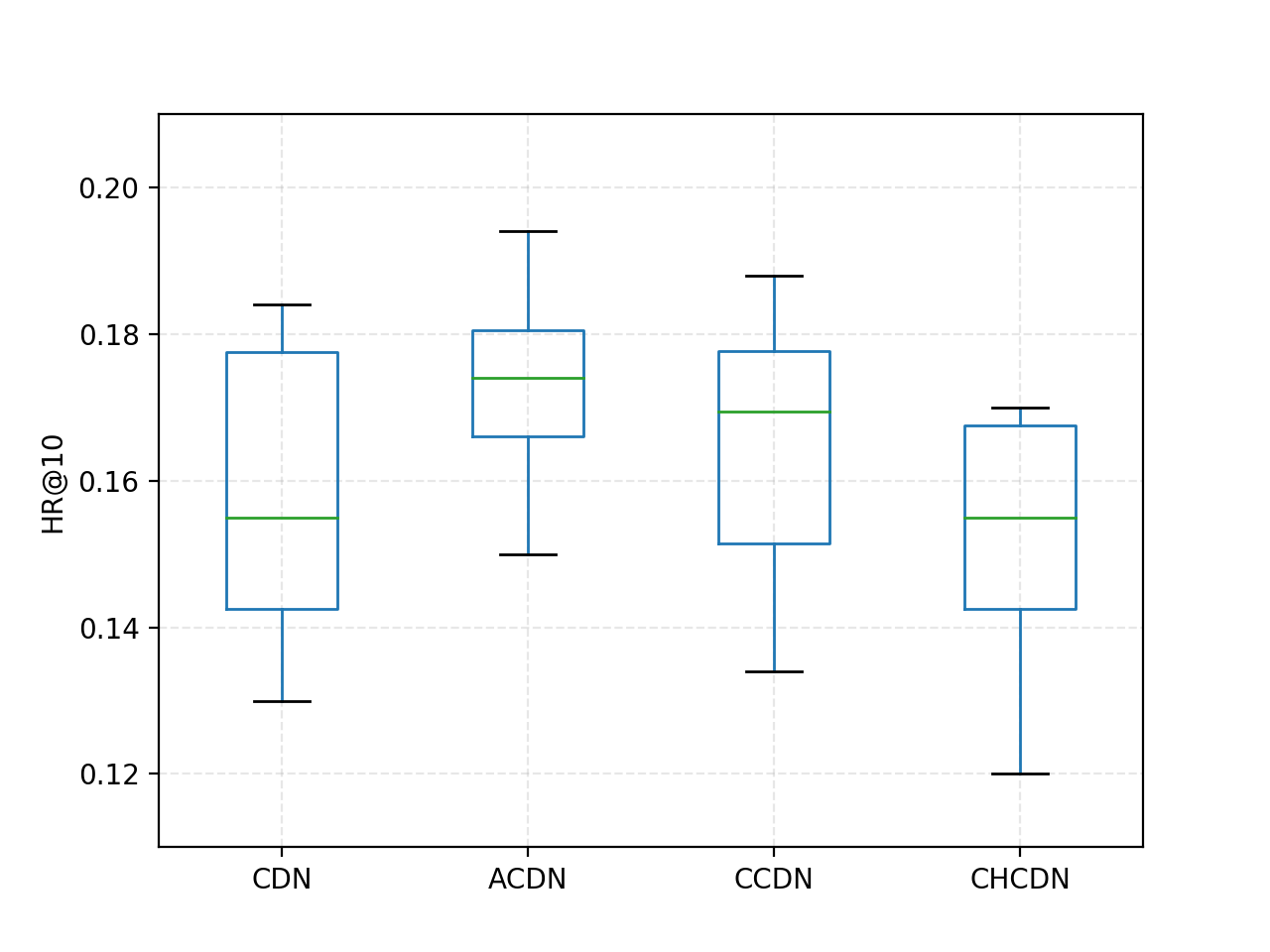}
	}
	\subfigure[Loss and Performance]{
		\label{subfig:epoch}
		\includegraphics[height=3.5cm,width=4cm]{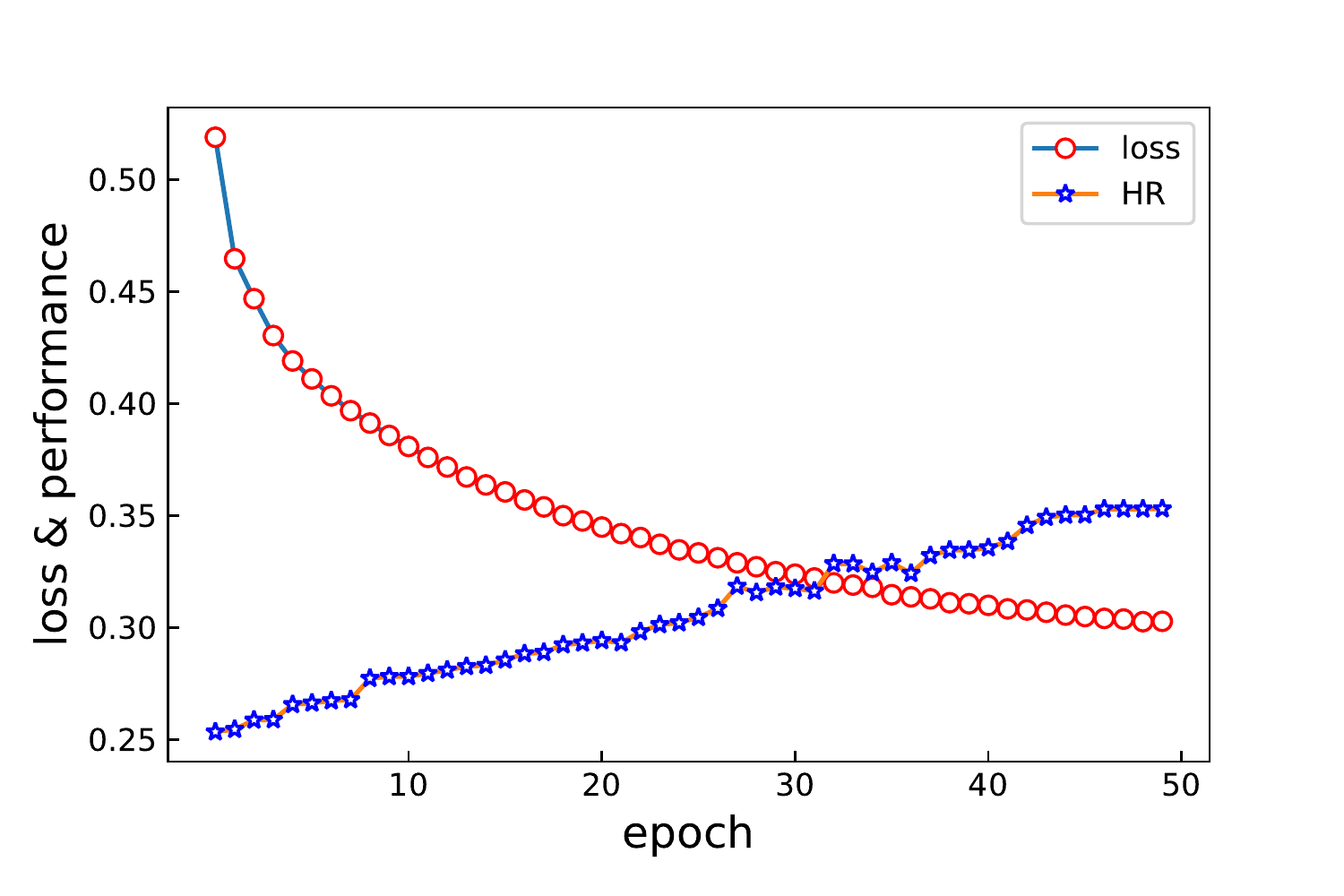}
	}
	\caption{(a) shows the comparison of various visual features. (b) is the analysis of optimization performance of our model.}
   \label{fig_param2} 
\end{figure}

In this subsection, we discuss the necessity of aesthetic features. We combine various widely used 
features in our basic model and compare the effect of each type of features by constructing models:
\begin{itemize}
	\item\textbf{CDN}: Removing the aesthetic features from our proposed model.
	\item\textbf{CHCDN}: Replacing the aesthetic features with color histograms of our model.
	\item\textbf{CCDN}: Replacing the aesthetic features with CNN features of our model.
\end{itemize}
Figure \ref{subfig:visual} shows the distribution of 10 maximum at HR@10 on Dataset 1 during 
40 iterations. We can observe that CHCDN performs the worst since the low-level features are too crude and unilateral, 
and can provide very limited information about consumers' aesthetic preference for cross-domain. 

Our model ACDN, with aesthetic information, performs the best, though CNN features also contain some aesthetic information (like color, texture, etc.).
It is far from a comprehensive description, which can be provided by the aesthetic features on account of the abundant raw aesthetic features inputted and training for knowledge transfer for cross-domain recommendation.
CNN features can perform better than aesthetic features in a single domain \cite{yu2018aesthetic}, but experiments demonstrate the effectiveness of the aesthetic features in cross-domain recommendation. 
This phenomenon proves our assumption that a user's aesthetic preference is domain independent and can be used as a bridge between domains for knowledge transfer.
\subsection{Impact of Hype-Parameters (RQ3)}
\subsubsection{Impact of $l_1$-norm Regularization}
Figure \ref{sparesize} shows the impact of $l_1$-norm regularization on the entries $h_{ij}$ of $\boldsymbol{H}^l$ in Eq.\ref{H} .
ACDN-$l_1$ is that we remove the $l_1$-norm regularization from our model. From the experimental results, we can observe that ACDN performs better than ACDN-$l_1$ on both datasets, which demonstrates the effectiveness of enforcing the sparse structure ($l_1$-norm regularization) on the cross transfer matrices. The  $l_1$-norm regularization can control knowledge transfer between source domain and target domain. In other words, with $l_1$-norm regularization, our model can utilize the cross transfer matrices to select representations adaptively to transfer for cross-domain recommendation.

\subsubsection{Sensitivity Analysis of  $\lambda$}
\begin{figure}
\setlength{\belowcaptionskip}{-0.6cm} 
   \centering
	\subfigure[Performance on Dataset 1]{
      \label{subfig:a}
		\includegraphics[height=3.5cm,width=4cm]{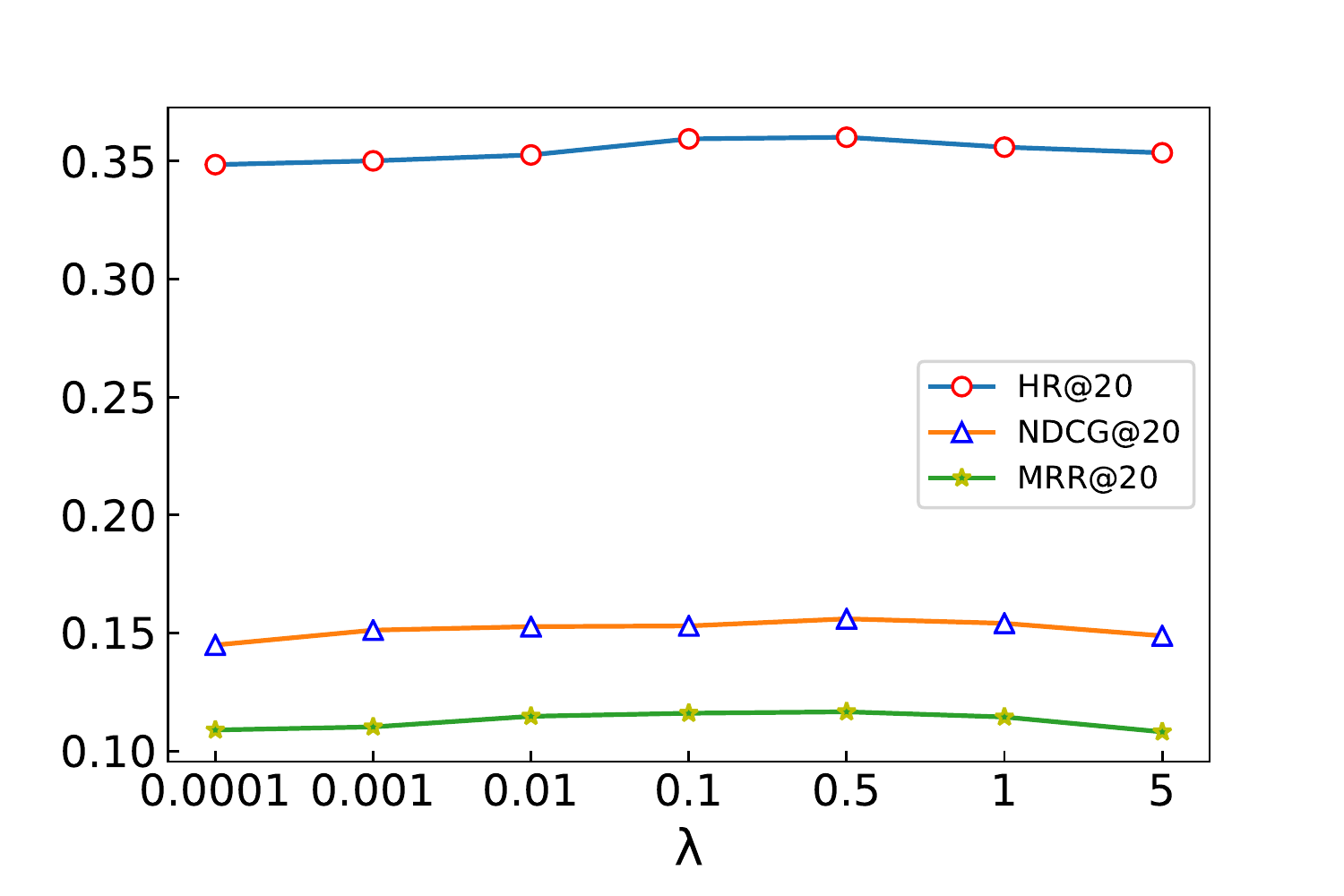}
	}
	\subfigure[Performance on Dataset 2]{
		\label{subfig:b}
		\includegraphics[height=3.5cm,width=4cm]{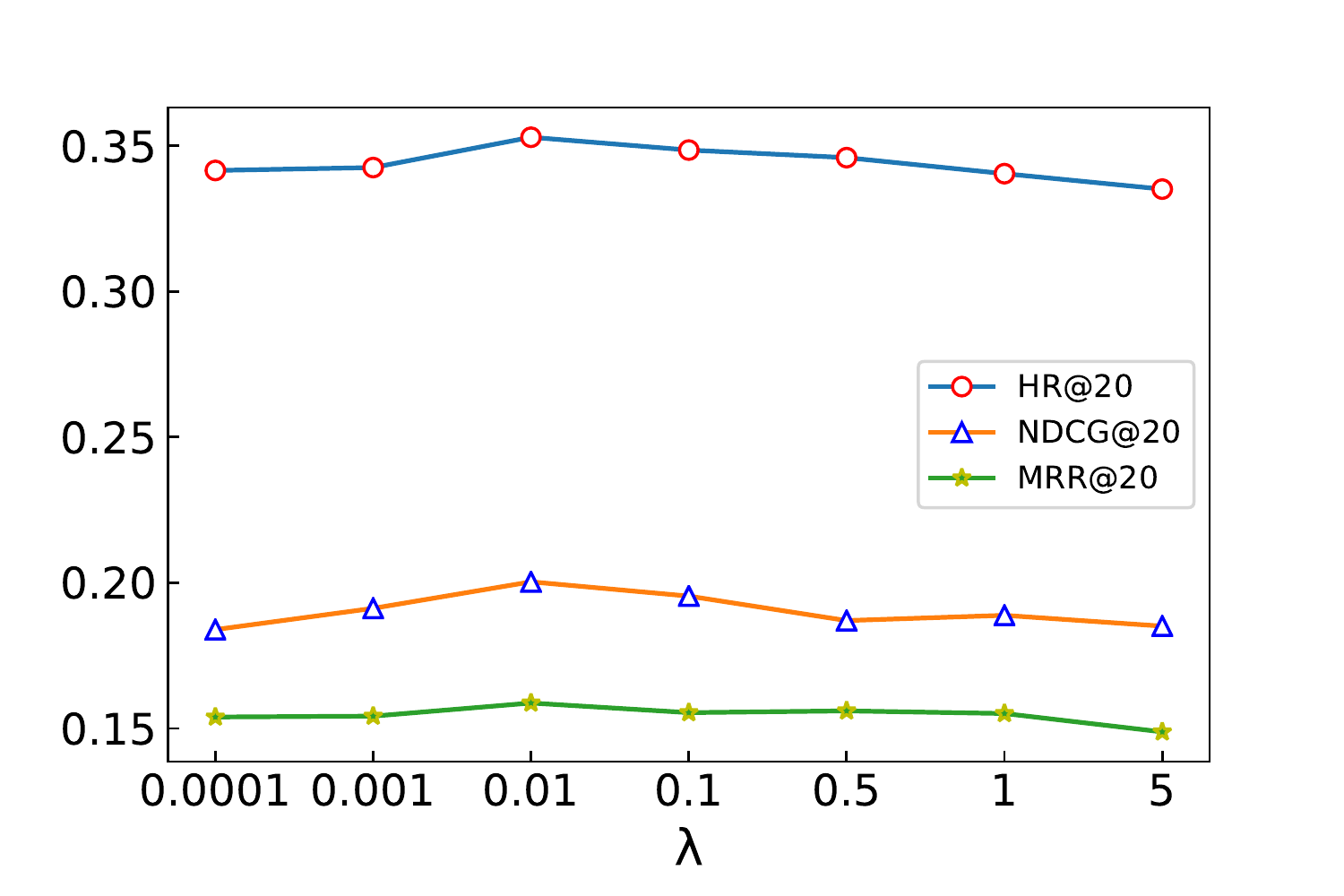}
	}
	\caption{Sensitivity Analysis of $\lambda$}
   \label{fig_param} 
\end{figure}

From the above analysis of impact of $l_1$-norm regularization on cross transfer
matrices, we can see that the $l_1$-norm regularization is crucial to our model. But how to set the appropriate penalty parameter $\lambda$ of $l_1$-norm regularization?
We will analyze the sensitivity of the penalty parameter $\lambda$ of $l_1$-norm regularization and we optimize the performance of our model varying with $\lambda \in \{0.001, 0.01, 0.1, 0.5, 1, 5\}$. 
As is shown in Figure \ref{fig_param}, our model achieves the best performance with setting $\lambda$=0.5 on Dataset 1,  while it achieves best performance with 
setting $\lambda$ = 0.01 on Dataset 2. It is possible that the two datasets have different distribution of information. Thus, setting appropriate
sparse penalty parameter under different background can improve the performance of our model.

\subsubsection{Optimization Performance}
We analyze the optimization performance of our model varying with training epochs. 
Figure \ref{subfig:epoch} shows the training loss 
and NDCG@20 test performance on dataset 2 (HR and MRR have similar trends) varying with each optimization iteration. 
We can observe that with more iterations, the training loss gradually decreases and the recommendation performance is improved accordingly. 
The most effective updates are occurred in the first 30 iterations, and its performance gradually improves until 40 iterations. With more iterations, Our model is relatively stable.

%% file: conclusion.tex
\section{Conclusions}\label{conclusion}
In this paper, a new deep Aesthetic preference Cross-Domain network (ACDN) was introduced to transfer users' aesthetic preferences across different domains to enhance the recommendation performance.
Specifically, we proposed a deep cross-domain recommendation network incorporated with aesthetic preferences, which enabled dual knowledge transfer across domains by introducing cross transfer unit from one base network to another.
Our work improved existing cross-domain recommendation research in two ways: 
(i) We leveraged novel aesthetic features for cross-domain recommendation to capture users' domain independent aesthetic preference; and
(ii) We proposed a new cross-domain recommendation algorithm for better modeling an individual's propensity from the aesthetic perspective, in which the aesthetic preference of each individual is shared for knowledge transfer across different domains to alleviate the data sparsity problem.
Using the Amazon dataset across three domains, we evaluated the effectiveness of our proposed approach against various baseline methods.
Experimental results showed that: 
(i) The aesthetic features were effective in cross-domain recommendation. This further demonstrated that users' aesthetic preference is domain independent.
(ii) We found that deep/transfer
models were superior to shadow/non-transfer methods, and incorporating aesthetic features into cross-domain recommendation could further improve the accuracy of recommendation.
(iii) Dual knowledge transfer across domains by introducing cross connections
from one base network to another can let
them benefit from each other, which is superior to the knowledge transfer in one direction.